\documentclass[11pt]{article}

\usepackage[latin1]{inputenc}
\usepackage{epsfig,graphicx}
\usepackage{latexsym}

\parskip =\medskipamount
\thicklines
\def\be{\begin{equation}}
\def\bea{\begin{eqnarray}}
\def\ee{\end{equation}}
\def\eea{\end{eqnarray}}

\newcounter{fig}

\def\d{\mbox{d}}

\def\fp{\displaystyle}
\def\ba{\begin{array}{rcl}}
\def\ea{\end{array}}

\begin{document}

\title{ Numerical studies of planar closed random walks. }

\author{ Jean Desbois and St\'ephane Ouvry      }

\maketitle	

{\small
\noindent
Laboratoire de Physique Th\'eorique et Mod\`eles Statistiques.
Universit\'e Paris-Sud, B\^at. 100, F-91405 Orsay Cedex, France.

}

\begin{abstract}

 Lattice numerical simulations for planar closed
 random walks and their winding sectors  are presented.
 The frontiers
 of the random walks and of their
 winding sectors  have a Hausdorff dimension $d_H=4/3$. However, when
 properly defined by taking into
 account the inner 0-winding sectors, the frontiers
 of the random walks have a 
 Hausdorff  dimension  $d_H\approx 1.77$.

\end{abstract}

\vskip1cm

\section{Introduction}\label{sec1}

Important progresses have been made in the past few years  for the
determination of critical exponents of planar Brownian curves based on a
family of conformally invariant stochastic processes, the stochastic Loewner
evolution (SLE) \cite{SLE}. It has been  proved \cite{Hausdorff} that
the external frontier has a Hausdorff
dimension  $d_H=4/3$, a confirmation of a numerical conjecture by
Mandelbrot \cite{Mandel}.  The  frontier of the Brownian curve is here defined
as the set of points, i.e the part of the
 curve,  where one stops  when arriving from infinity and   meeting it
 for the  first time (the hull).
 This definition is  geometrical,  only the geometrical outer spreading of the curve does  matter. 

It is well known that random walks with $N$ steps approximate well planar
Brownian curves of length $t$ when the number $N$  becomes large. The correspondance between $N$  
and $t$  is $Na^2=2t$, where $a$ is the lattice spacing  (it will be set  to $1$ in the following). 
However, this statement has to be refined when winding properties
are considered.  For instance, the probability
 density of the angle
 $\Theta$ wound around the origin by an open Brownian curve is
 given, in the asymptotic regime, by  Spitzer's law \cite{Spit} 
$\fp P \left( x={2\Theta}/{\ln t} \right)={1}/({\pi}({1+x^2}))$. 
On the other hand, for a random walk on a
 two dimensional (2d) square lattice, the probability density  is given by   Belisle's law \cite{Beli}
$\fp P \left( x={2\Theta}/{ \ln N } \right)={1}/({2}
{\cosh(\pi x/2) })$. Note that  all the moments are defined in the latter case whereas  
no moment exists in the former case, a fact  which can be traced back
to short-distance properties
 of Brownian motion, which are  ignored when the lattice spacing is non zero.

In this paper, we  adress numerically  some
 properties of closed random walks on a 2d square lattice
 and discuss whether these results can also hold  for Brownian curves. 
More precisely, we are concerned by fractal properties of
the random walks and of their winding sectors.

For a closed Brownian curve of length $t$,  a $n$-winding
sector inside the curve is defined as a connected set of points that have been encircled $n$ times
by the curve (more precisely $n$ is the number of anticlockwise minus
clockwise encirclings).
\begin{figure}
\begin{center}
\includegraphics[scale=.20,angle=0]{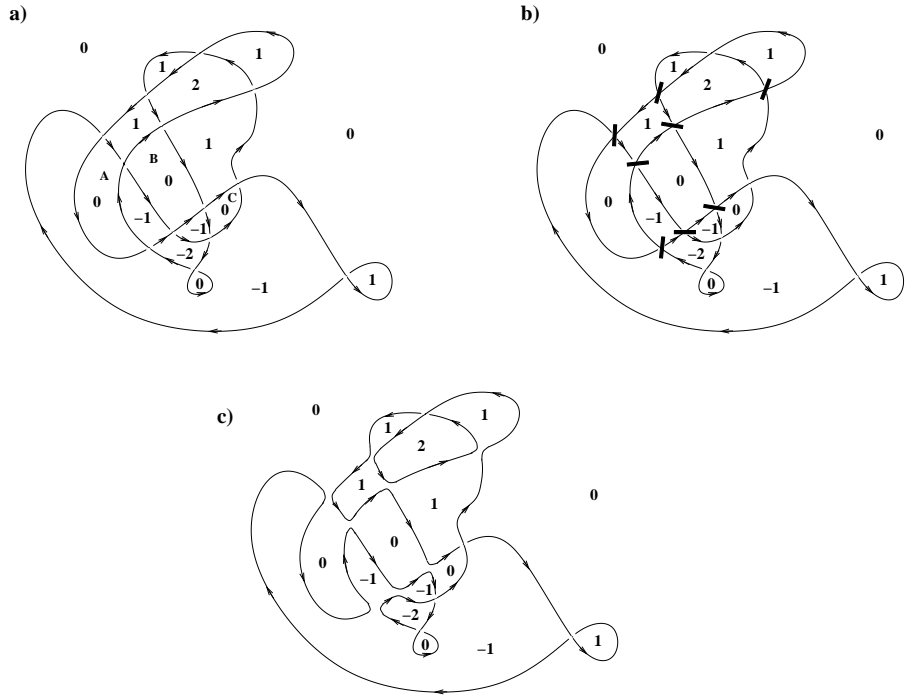}\label{fig1}
\caption{ a) a closed curve with its winding sectors; 
b) the cuts required to open the fjord (see section \ref{sec2}); c) the  closed curve  with
 the fjord  opened. a) and c) represent two possibe time histories of the same curve.}
\end{center}
\end{figure}
As an example, consider the  curve of figure 1 a):  
leaving aside the infinite outside $(n=0)$-winding sector, which  we
 call the "sea", this curve  encloses five
 ($n=1$)-winding sectors, three ($n=-1$)-winding sectors,
 one   
($n=2$)-winding sector,  one  ($n=-2$)-winding sector, and four
$(n=0)$-winding sectors.  Among them, three -marked as
  A, B and  C-  can be connected between themselves and to the
  outside ''sea'' by simple cuts as indicated in
 figure 1 b). For this reason, they  constitute what is called a
 "fjord". The last $0$-winding sector does not share this property,
 lying deep inside the curve. It is  called  a ''lake''. 
 
Let us denote by $S_n$\footnote{In the random magnetic  impurity model \cite{Random}, a crucial role
is played by
 the joint probability distribution of  the random variables
 $S=(2/t)\sum_nS_n \sin^2(\pi\alpha n)$ and $A= (1/t)\sum_nS_n
 \sin(2\pi\alpha n)$,
 where $\alpha$ is the Aharonov-Bohm flux (in unit of the quantum of
 flux)
carried by the magnetic impurities.} the
 arithmetic area of all the  $n$-winding sectors of a  Brownian
 curve of length $t$,  and   by $S_0$ the  arithmetic area of all the 0-winding 
sectors inside the curve, i.e.  the fjords and  lakes.
Scaling properties for Brownian curves imply that the random variable $S_n$ scales like
$t$. Its expectation value $<S_n>$ on the set of all
closed curves of length $t$ has been computed by path integral methods  \cite{Sn}.
As long as $n \ne 0$,   $<S_n> ={t}/{( 2 \pi n^2 )}$. On the
other hand, when $n=0$, the 0-winding sectors
 arithmetic area   diverges because, in this approach, the
 outside sea is necessarily
 taken into account. 
It was 
later shown by  Werner    in his
thesis \cite{Thesis} that, for $n$ sufficiently large, $n^2S_n\to
<n^2S_n>=t/(2\pi)$. 
Recently, using SLE methods \cite{Jeunes}, the 
  expected value of the  total
 arithmetic area $<S>=\sum_{n=-\infty}^{n=\infty}<S_n>$ of the
 curve has been obtained to be  $<S>= t \pi/5$. It implies for the
 arithmetic area of the 0-winding 
sectors inside the curve $<S_0>=t\pi/30$, a rather  striking result,
bearing in mind that the lakes and fjords arithmetic 
areas   were up to now out of reach, at least by methods inspired from physics.

\begin{figure}
\begin{center}
\includegraphics[scale=.3,angle=0]{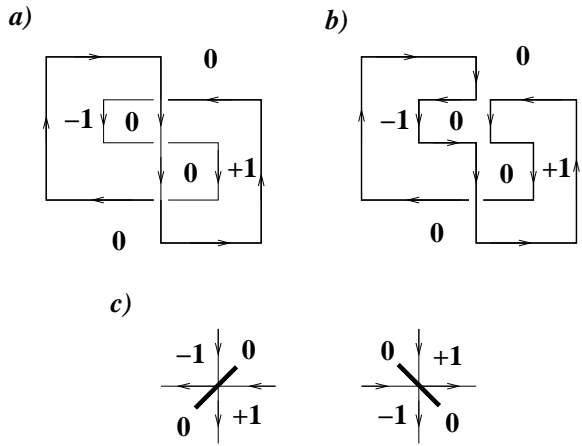}\label{fig2}
\caption{ a) a closed walk with its winding sectors; 
 b) the same walk with an oriented frontier (the fjord is opened; see
section \ref{sec2}); c) the two cuts needed to go from a) to b).  a) and b)  are two possible
 time histories of the same walk.}
\end{center}
\end{figure}

 For closed random walks, the winding sectors can be defined in a similar way to those of  Brownian curves:
each elementary cell of the  lattice is labelled by its winding
number $n$; two adjacent cells with the same $n$ are connected 
 if their commun edge is not spanned by the walker. A winding sector
 is a set of connected cells.   
 For instance, the random walk in figure 2 a) encloses one (n=1)-sector, one
(n=-1)-sector and a fjord  constituted of the two $0$-winding
sectors.

\begin{figure}
\begin{center}
\includegraphics[scale=.35,angle=0]{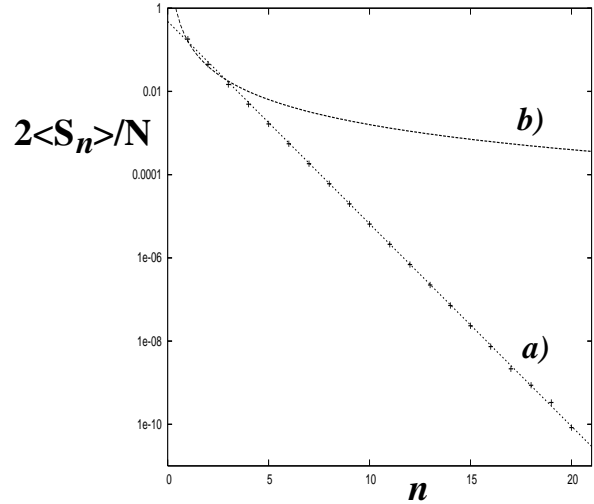}\label{fig3}
\caption{ the  average arithmetic area $<S_n>$ of all the $n$-winding sectors 
  as a function of  $n$;
 a) numerical simulations for random walks-- the straight line is an
  exponential fit; b) the analytical result for Brownian curves.  }
\end{center}
\end{figure}

   In figure 3 a),  the  average arithmetic area  $2<S_n>/N$ of  all the $n$-winding sectors inside the random walk 
 (log scale) is plotted as a function of the winding number $n\ge 0$, for
 10000 closed random walks of $N=6000000$ steps. Clearly, 
 the   behavior is exponential as soon as $n \ge 3$. This is to be compared
  to the Brownian  path integral result $<S_n>/t =1/(2\pi n^2)$  
(dashed curve in figure 3 b)). Moreover, for $n=0$, one obtains $2<S_0>/N \approx 0.13$ 
  in place  of the SLE result  $<S_0>/t= \pi /30 $ for Brownian curves.
 
 In figure
  4, on the other hand,   the average total arithmetic area $2 <S>/N$ 
  is plotted as a function of the number of steps $N$,  
  from $N=4000$ to $N=8000000$. It is manifest that as $N$ increases,
$2 <S>/N$ is closer and closer to the SLE result  
$ <S>/t=\pi /5$ for Brownian curves. Thus, one 
 concludes  that  the numerical lattice simulations for the $<S_n>$'s  do not yield
 the Brownian  path integral result $t/(2\pi n^2)$, but
 their total sum  correctly reproduces the SLE result.

\begin{figure}\label{fig4}
\begin{center}
\includegraphics[scale=.3,angle=0]{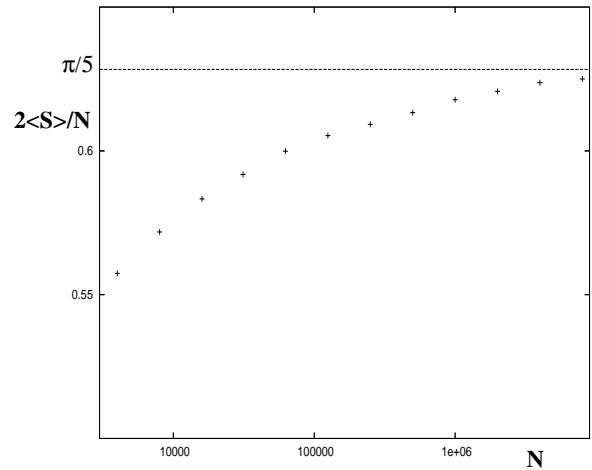}
\caption{ the average  arithmetic area enclosed by random walks as a function
  of the number of steps $N$; the $\pi/5$ horizontal line is the 
  analytical result for  Brownian curves.}
\end{center}
\end{figure}
 This means that, quite generally and as expected, global properties of 
  random walks do indeed  hold for Brownian curves. 
A good agreement is also obtained for example  while comparing the radii of
 gyration of random walks to those of Brownian curves, as we will see in Section 2.
However, when  
details (for instance the properties of
winding-sectors taken separately) are considered, the situation is
less favorable.
 In that case, it is  not a priori clear if and how numerical
 simulations on random walks
 give non ambiguous  informations on Brownian curves, as already
 seen for 
 Spitzer and Belisle's laws.

\section{Frontiers of  closed  random walks}\label{sec2}

Let us turn to the fractal properties of the external frontiers of  random walks.
When estimating the
scaling and the Hausdorff dimension of the perimeter of its frontier, a
characteristic length for the random walk is needed.
 A natural candidate 
is its radius of gyration, for which two  definitions  are possible. 
Consider first the successive positions
$\overrightarrow{r_i}, \; i=1,2,...,N,$  of the  random walker
and define the radius of gyration $r_g^{(1)}$ as

\bea
    \left[r_g^{(1)}\right]^2 & = & \frac{1}{N} \sum_{i=1}^N\left(
      \overrightarrow{r_i} -  \overrightarrow{G}^{(1)}   \right)^2 \\
\overrightarrow{G}^{(1)}  & = & \frac{1}{N} \sum_{i=1}^N \overrightarrow{r_i}
\eea

Alternatively, consider the
 part $\cal D$  of the plane enclosed by the random walk as an
 homogeneous solid  of area $S$ and compute the  radius of gyration
  $r_g^{(2)}$ of this solid

\bea
    \left[ r_g^{(2)} \right]^2 & = & \frac{1}{S} \int \int_{\cal{D}} 
\left( \overrightarrow{r} -  \overrightarrow{G}^{(2)}   \right)^2 \d S\\
\overrightarrow{G}^{(2)}  & = & \frac{1}{S } \int \int_{\cal{D}} \overrightarrow{r} \d S
\eea 

\begin{figure}\label{fig5}
\begin{center}
\includegraphics[scale=.3,angle=0]{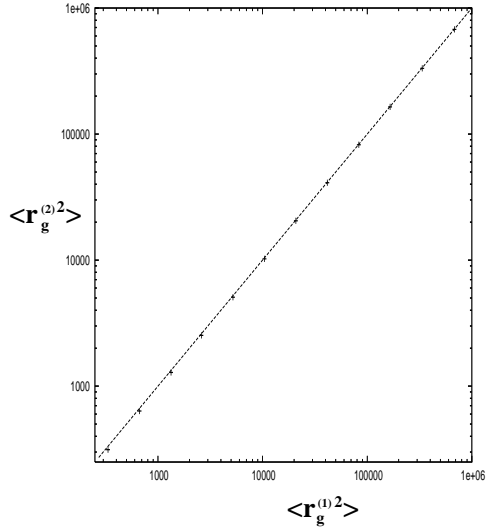}
\caption{ Comparison of the two definitions of the radius of gyration of
  closed random walks; the straight line is $<r_g^{(1) 2}> = <r_g^{(2) 2}>$.}
\end{center}
\end{figure}

Numerical simulations  of $r_g^{(1)}$ and  $r_g^{(2)}$ are presented in
figure 5 ($N$ runs from $N=5000$ to $N=8000000$, the straight line is the
diagonal). Clearly, both definitions lead to very close
numerical estimates so that one can use indifferently the
radius of gyration 
  $ r_g =(r_g^{(1)}\simeq r_g^{(2)})$
 as a characteristic length for the random walk. 

Incidentally, and as a further check of the  correspondance 
between random walks and Brownian curves as far as their global characteristics are concerned,
 the average total arithmetic  area $<S>$ as a function of $<r_g^2> $  is plotted  in figure
 6. The  $r_g^{(1)}$ distribution, which has already been  computed analytically for
closed Brownian curves, again 
 by means of path integral methods \cite{FD},  gives 
$\fp  < \left(  r_g^{(1)} \right)^2 > = {t}/{6}$.
Using this input, one should get 
\be \label{fine}
  <S> = \frac{\pi}{5} t = \frac{6 \pi}{5} <r_g^2> \approx 3.77   <r_g^2>       
\ee
which is indeed the straight line plotted in figure 6.

Interestingly enough,   (\ref{fine}) is related to 
  random
   walks   anisotropies- clearly, for an homogeneous isotrope disk, one should have  $S=2\pi
  r_g^2$.
The ratio ${\lambda _+}/{\lambda _-}$, where $\lambda _+ > \lambda _-$ are the eigenvalues of the inertia tensor,
can also be used to characterize the anisotropy. For closed Brownian curves, it  has been known for some time \cite{FD} 
that 
\be\label{asph1}
  \frac{< \lambda _+ >}{< \lambda _- >} \; = \; \frac{ 1+ 3 \ln 2 - \pi/2 }{ 1- 3
    \ln 2 + \pi/2 } \; \approx \; 3.07
 \ee
pointing to an elongated shape. We will come back to this point in Section \ref{sec3}.

\begin{figure}\label{fig6}
\begin{center}
\includegraphics[scale=.3,angle=0]{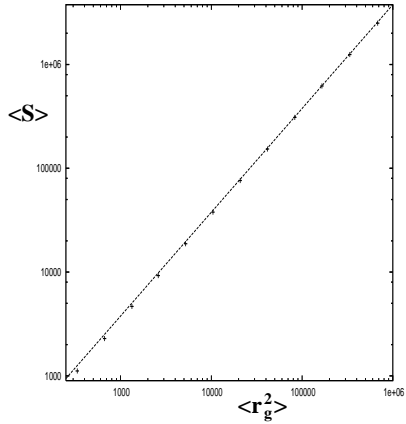}
\caption{ the arithmetic area as a function of the square of the  radius of gyration; 
the straight line is the fit $ <S> = 3.77 <r_g^2>$.}
\end{center}
\end{figure}

Once the characteristic length $r_g$ of a random walk  has been identified,
one can turn to investigate the scaling 
 of the perimeter
of its external  frontier with respect to this length. The external frontier is geometrically defined,  accordingly to the external frontier of a Brownian curve, as the set of points, i.e. the part of the walk, obtained by arriving from infinity and stopping when one meets for the first time the walk.  
In figure 7 a), the scaling  is displayed for closed random walks with a
number of steps from $N=16000$ to $N=8000000$. The straight line
shows that the external frontier scales like $\fp r_g^{(4/3)}$ leading  to the
same  Hausdorff dimension  $d_H=4/3$ as the one obtained via SLE   for Brownian curves. 
Therefore, as expected for a global characteristics, the scaling of the perimeter of the random walk correctly reproduces the Brownian curve Hausdorff dimension.

\begin{figure}\label{fig7}
\begin{center}
\includegraphics[scale=.4,angle=0]{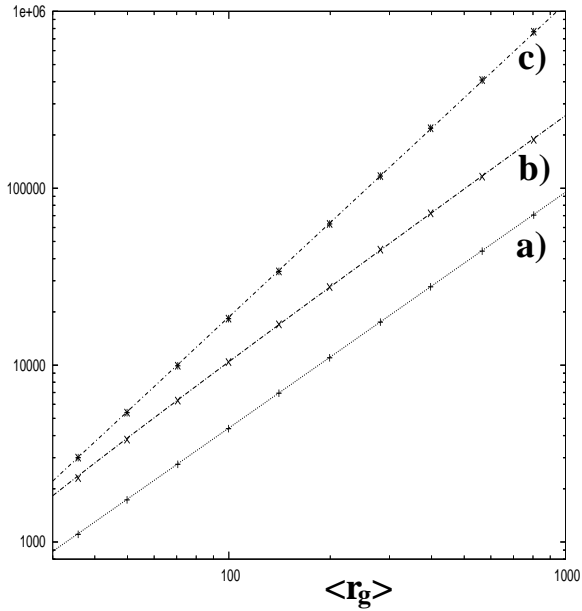}
\caption{ numerical results for  the frontiers perimeters of closed random walks
 (the number of steps varying from $N=16000$ to $N=8000000$) 
 as functions of the
  radius of gyration; a) external (geometric) frontier;  b) external+fjords (oriented) frontier;
 c)  external+fjords+lakes  (disconnected) frontier; the slopes of the straight
 lines are : a) 4/3 , c) $1.77\approx 7/4$. See the text for b) (logarithmic
  corrections are included).  }
\end{center}
\end{figure}

\section{Oriented frontiers of closed random walks}

Up to now, only the geometrical outer spreading of the random walks or of the Brownian curves   has
been taken into account to define their  external frontiers.   A  consequence is that there is no way to   span the frontiers defined in this way with a
 coherent orientation, as can be seen for example in figure 1 and figure 2. 
The impossibility to have a coherent orientation might incline to reconsider
 the definition of the frontiers. To this end, one should  pay a particular
attention to the
 possible time histories of a given random walk or  Brownian curve. 

A  random walk   on the lattice is nothing but  a set   of   oriented edges  spanned
with Kirchoff conservation
 laws at the vertices.    Consider first the  random walk of  figure 2. The
 two 0-winding sectors 
 happen to be  next to the  outside sea. They constitute a  fjord 
 which can be   opened to
 the outside by two simple cuts at particular    intersections. In figure 2 a) and
 figure 2 b) are shown   two
 possible time histories of the  random walk (by convention when a random
 walk  intersects itself at
 a later  time  it goes under its first trace). On the one hand, 
 the  time history  of figure 2 a) is such that   both  the $0$-winding
 sectors are  inside the walk
 (the fjord is closed) and  its external frontier  coincides indeed
 with its geometrical outer spreading
(i.e. the part  of the walk where one
 stops when arriving from infinity and
 meeting for the  first
 time the  walker). As already stressed,   there is   no  coherent orientation on such a 
 frontier.  On the other
 hand,  the time history of figure 2 b) is such that both the 0-winding
 sectors of the walk
 are  now connected between themselves and to the sea (the fjord  is  opened).  A  
  frontier  can then be defined  by following  the time history in figure 2 b) and, while doing
 so, by selecting
 the part of the walk   which  always has on one of its side the
 $0$-winding sector
 (the outside) and, necessarily, on the other side a ($n\ne
 0$)-winding sector\footnote{In this simple case the frontier is the random walk itself.}.
  It is manifest that this frontier is,    by construction, spanned  with a  coherent
 orientation. Contrarily to the external geometrical frontier, it 
explores the inside of the walk while spanning  the opened fjord.
 Rephrased in terms of cuts at intersections,  the two cuts depicted 
 in figure 2 c) are   used to transform the  walk 
 of figure 2 a) into the  walk of figure 2 b). Note that in
 this cutting process with an even number of cuts -here 2-, one has  obtained a single random walk 
(it would not  be the case if, for example,  only one cut would have been
used: two disconnected walks would then have been obtained).

Let us take advantage of what has been said but now for  Brownian curves, 
 for example the  curve of figure 1.   Consider  the
 time history in
 figure 1 a): the external frontier  is the geometrical one,
 as well as 
 the frontiers of the $n$-winding sectors inside, which are all
 disconnected.
 The three 0-winding sectors which could be  connected to the outside
 by
 simple cuts  are inside the curve (closed fjord). However,
eight  cuts, as shown in figure 1 b), are sufficient to connect
between themselves
 these 0-winding sectors and to open to
 the outside  the  fjord. Also, these cuts  connect between themselves,
 when possible,  the winding sectors of same winding number
 (here respectively the four  $1$-winding sectors in the upper 
 part of the path  and the three $(-1)$-winding sectors in the
 lower part of the curve). One  arrives at the Brownian curve
 of figure 1 c)  whose  frontier is obtained  by
 following the time history in figure 1 c) and, while doing so,
  selecting  the  part of the curve which always has on one of its side 
 the $0$-winding sector (the outside) and on the other side
a ($\pm 1 $)-winding sector. This frontier has, by construction, a
coherent orientation\footnote {The frontiers of the $(n=1,2,-1,-2)$-winding sectors
   inside the curve  are obtained 
 accordingly  by  following the time history of the curve  and
 selecting   the  part
  which has always on one of its sides the $n$-winding sector  and on the other side a $(n\pm 1$)-winding sector.
By construction, these $n$-winding frontiers  have  a coherent orientation.}. 
 One also notes that the lake inside
 the $(-1)$-winding sector remains disconnected from the outside sea.

More generally, for any given Brownian curve, one can always find a
time history with
  an oriented frontier  consisting of  the   external + fjords frontier.
 Clearly, this oriented frontier is  more intricate than the usual one 
since it does excursions   inside the curve while  spanning the
opened fjords, which always happen
 to stretch from one side to another side of the curve.
  
One might wonder if these excursions around the fjords are sufficiently intricate  to increase the  Hausdorff dimension of the frontier, if such a dimension happens to  exist.
 Numerical simulations in figure 7 b) indicate that  the
  external+fjords (oriented) frontier perimeter scales like $\fp      r_g^{(4/3)}(\ln (r_g/8.5))^{0.205} $. Therefore, up to logarithmic
  corrections,  the Hausdorff
  dimension is essentially unchanged and  remains equal to  $4/3$.
However, when pushing further the logic of avoiding the $0$-winding sectors  by also removing the lakes inside the random walks, i.e. 
when considering the external+fjords+lakes (disconnected)  frontier,
 the numerics in figure 7 c)  indicate a
scaling with a  Hausdorff dimension   $d_H\approx 1.77 \approx
7/4$.  The increase of the dimension  is nothing but an indication of
the more and more intricate nature
of the frontier, which goes not only around the fjords, but also  deep
inside the curve around the lakes
 (see figure 8 for an illustration of the distribution of fjords and
 lakes  for a random walk of
   $N=1000000$ steps). Numerical simulations show that the ratio
   of the  lakes average area to the fjords average area increases
    from $\approx 0.4$ to $\approx 1.9$ when $N$ goes from 15000 to
    1000000, indicating that the lakes proliferate  deep inside the walk  when $N$ increases.
  
\begin{figure}
\begin{center}
\includegraphics[scale=.40,angle=0]{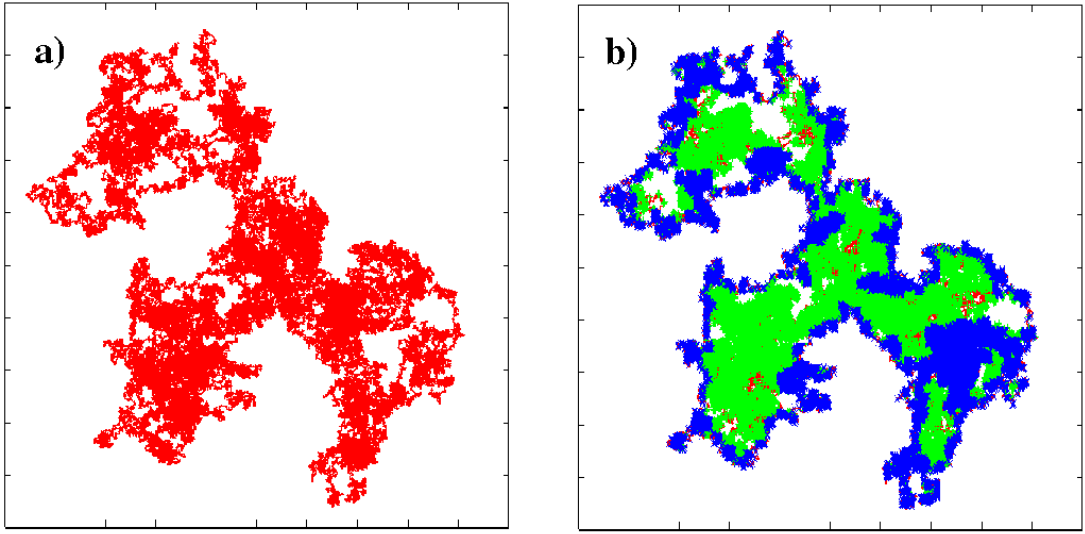}
\caption{ a) in red the closed random walk  with $N=1000000$; b) in
  blue the fjords, in green the lakes.
 The fjords remain in general on the boundary of the walk,
 whereas the lakes proliferate deep inside the walk.}
\end{center}
\end{figure}

Both Hausdorff dimensions $4/3$ and  $7/4$ are familiar in percolation
\cite{Du}. A sketch of a site percolation cluster on a 2d lattice is given in figure 9
a) and numerical simulations for lattice sizes  from
 100$\times$100, 
  200$\times$200, ..., up to 3200$\times$3200, are presented in figure 10.
\begin{figure}\label{fig9}
\begin{center}
\includegraphics[scale=.25,angle=0]{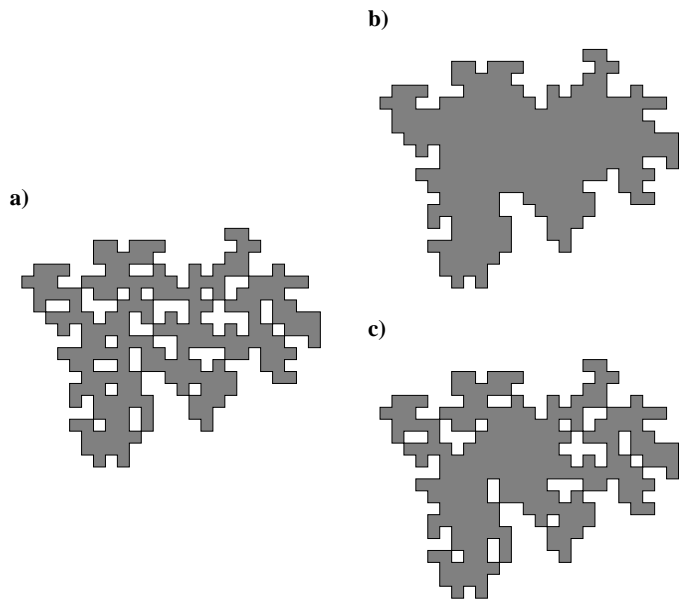}
\caption{ a) sketch of a site percolation cluster;  b) its  external frontier; 
 c) its external+fjords frontier.}
\end{center}
\end{figure}
\begin{figure}\label{fig10}
\begin{center}
\includegraphics[scale=.4,angle=0]{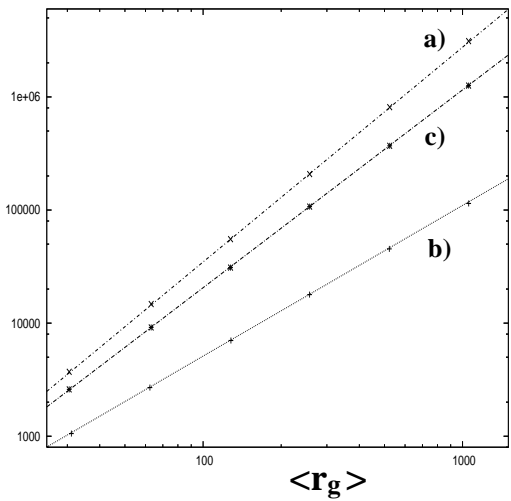}
\caption{ numerical results for site percolation  ($p_c\approx 0.593  $); the
size of the 2d square lattice runs from 100$\times$100 to 3200$\times$3200;
 in correspondance with figure 9: b) the external frontier perimeter 
 as a function of the cluster radius of gyration;
 c)   the external+fjords frontier;
 a)   the external+fjords+lakes frontier; the slopes of the fits are
 b) 4/3; c) 7/4; a) 1.9}
\end{center}
\end{figure}
On a 2d square lattice, at critical site percolation  occupation $p_c \approx  0.593 $, the 
  external frontier (see figure 9 b)) of   
a percolation cluster, defined accordingly to the external
frontier of  random walks as arriving 
 from infinity and
 stopping on the
 cluster, has
 a Hausdorff dimension $4/3$ (see figure 10 b)), as for random walks.
 In analogy with 
 random walks, if one removes
 inside the percolation cluster  the non percolating
 islands (see figure 9 c)) 
which can be connected to the outside   by  cuts at vertices (``fjords''),  the Hausdorff dimension of the external+fjords frontier becomes
 $7/4$ (see figure 10 c)).
 If one also  removes the
  non percolating  islands deep inside the percolation
 cluster (``lakes''), one arrives at a disconnected
 percolation external+fjords+lakes frontier (see figure 9 a))
 for which numerical simulations (see figure 10 a))   indicate a  Hausdorff  dimension 
 $d_H\simeq 1.9>7/4$. Percolation and closed random walks happen to 
have only in commun the scaling dimension $4/3$ for their 
external frontier,
 the  other dimensions being different ($7/4$ versus $4/3$, and
$\simeq 1.9$ versus $\simeq 1.77$). It is however  striking to note that  in  percolation   the fjords are sufficiently intricate to increase the Hausdorff dimension of the perimeter, whereas there are not so for  random walks.

\section{Frontiers of winding sectors}\label{sec3}

A $n$-winding sector has been defined in section
\ref{sec1} as a set of connected cells of same winding $n$ inside a random walk (see figure 11).
\begin{figure}\label{fig11}
\begin{center}
\includegraphics[scale=.45,angle=0]{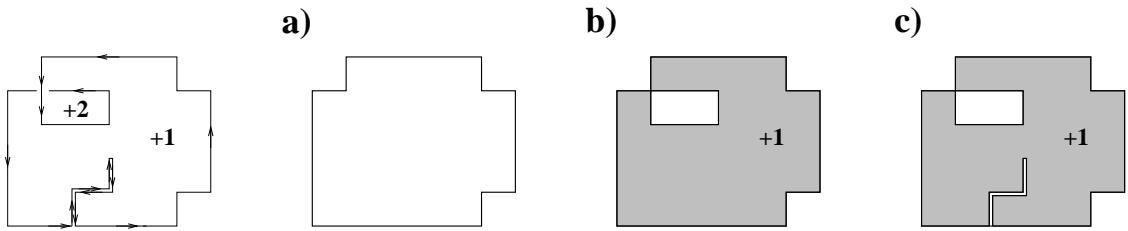}
\caption{  a (n=1)-winding sector; a) external (geometric) frontier; b) external+other windings frontier; c) external+other windings+ back-trackings  frontier.  }
\end{center}
\end{figure}
Numerical simulations for the $n=0,1,2,3$-winding sectors are presented in
figure 12 and in figure 13 
for $32000$ random walks of $N= 4000000$ steps. For each 
 walk, the radius of gyration, the enclosed arithmetic area, the
 anisotropy and the frontier are
 evaluated. 
 Insufficient  statistics makes it delicate to address  larger
 $n$ which would require an increase of $N$ leading to
 rapidly prohibitive CPU times.

In figure 12, one considers, for  each winding number, the set of all the corresponding winding sectors inside the random walks. 
  The average value $S$ of   the area enclosed by these winding
  sectors, plotted as a function of  the square of their   radius of gyration, leads  to the scaling 
    $S=3.77r_g^2$. This scaling is similar  to  the scaling of  the arithmetic area of the random walks thenselves,   plotted in figure 6. 
\begin{figure}\label{fig12}
\begin{center}
\includegraphics[scale=.35,angle=0]{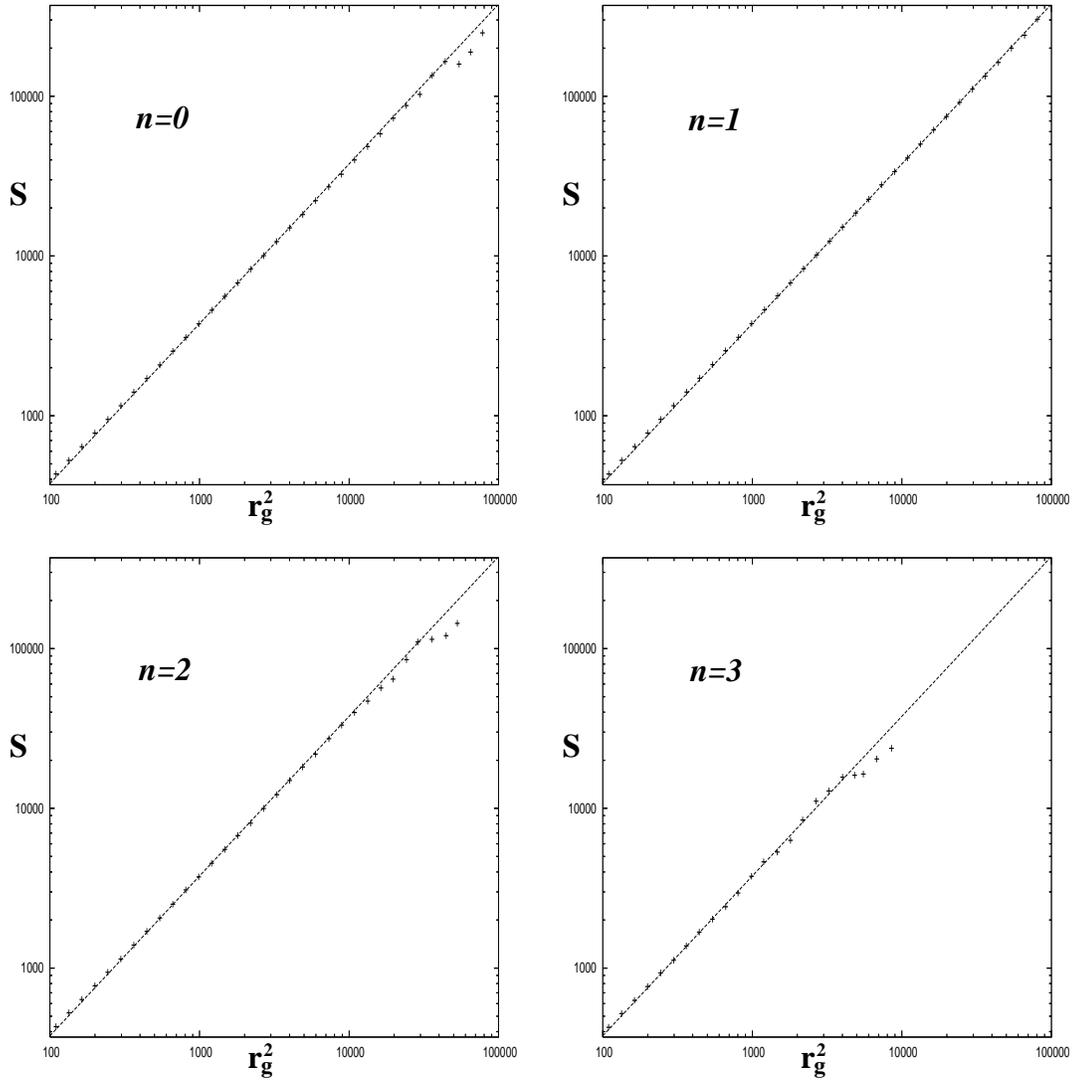}
\caption{ the arithmetic area, $S$, enclosed by the $n$-winding sectors ($n=0,1,2,3$)
 as a function of their radius of gyration; the straight lines are the
 fits $S=3.77r_g^2$.}
\end{center}
\end{figure}
 This  result is somehow surprising
   since a winding sector by itself cannot be
considered as a closed random walk.  Nevertheless, 
anisotropy properties of the winding sectors do confirm this
similarity
 between the random walks and their winding sectors. Numerical
 simulations for the ratio $\lambda _+/\lambda _-$  give for
the  $n=0$-winding sectors a   value in the range 
  2.9-3.1 and for the $n=1,2,3$-winding sectors  values in the range 2.9-3.2,
 which slightly increase with $n$. Again,  these results
   are close to  $<\lambda _+>/<\lambda _- > \approx 3.07$  of Section
   \ref{sec2}  for the random walks.

In figure 13, the scaling of  the winding
sector  frontiers perimeters is considered.   Let us first look
at the perimeter of a winding sector
  external  (geometric) frontier (see figure 11 a)).
  The  scaling
  (straight line a))   is  $r_g^{1.33}$, leading to  a
 Hausdorff dimension $d_H\approx 4/3$ which is close to the dimension   of 
   the random walk  frontiers.

\begin{figure}\label{fig13}
\begin{center}
\includegraphics[scale=.3,angle=0]{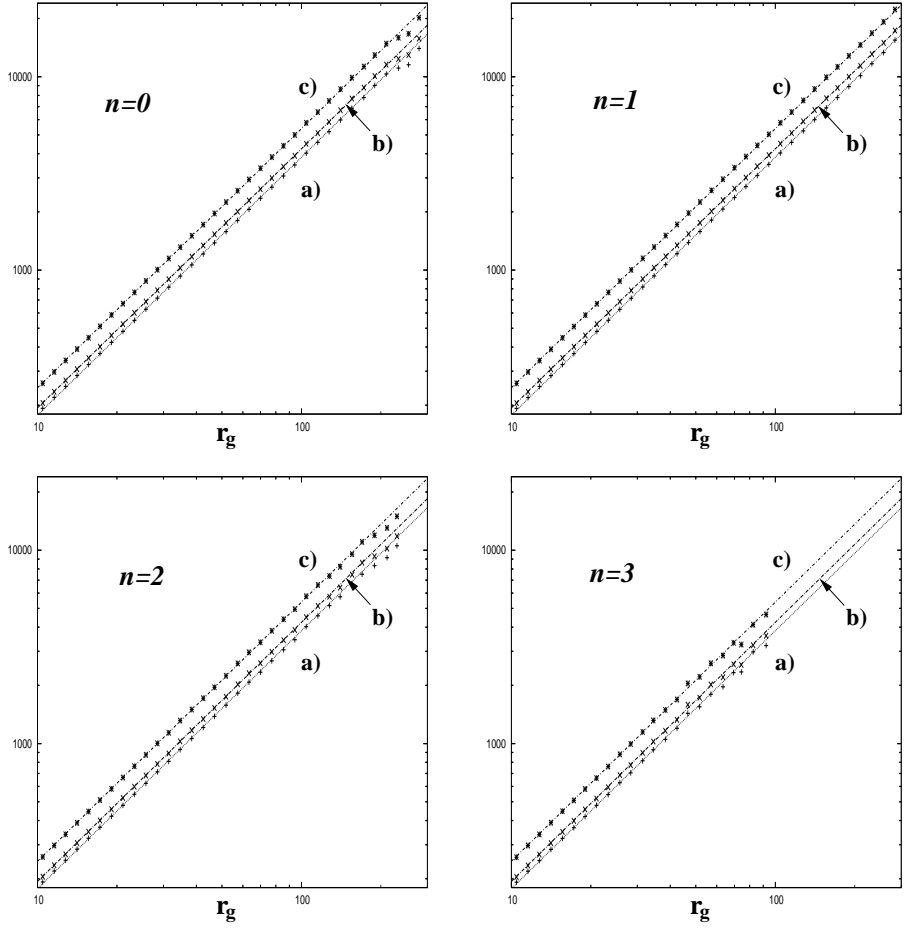}
\caption{numerical results for  the frontiers perimeters of the $n$-winding sectors ($n=0,1,2,3$);
 a)  external  frontier; b)
external+other windings frontier; c) external+other windings+back-trackings  frontier; the slopes of the straight
 lines are  a) 1.33 , b) 1.34 and  c) 1.34 .  }
\end{center}
\end{figure}

 Now, a given winding sector can enclose other winding sectors: one
can then define the  frontier of the winding sector as  its external frontier plus
the frontiers of the enclosed sectors (see figure 11 b).   
  This is the frontier  seen from the inside of the $n$-winding
 sector  and avoiding the cells belonging to the other sectors. 
The  scaling 
 (straight line b)) is  $r_g^{ 1.34 }$,
  which again leads    to the  Hausdorff dimension $d_H \approx 4/3$. 

A random walk  may also exhibit many back-trackings :
 one can  define the  frontier (see figure 11 c)) as  the external frontier + the other windings +
  the edges inside the sector which have been spanned by the walker in both directions. 
 The scaling (straight line c)) is still $r_g^{1.34}$, leading again to the  dimension $d_H\approx 4/3$.
 It follows that whatever  the definition of the frontier, 
 the scaling  is robust with a Hausdorff dimension remaining  approximatively    
  $d_H\approx 4/3$.    

As a way of testing this robustness, one could relax a little the definition of a winding sector by considering  that two
  adjacent cells belong to the same sector provided they have the same
  winding number, regardless   if their common edges  are spanned or not by the walker (see an example  in
  figure 14). 
\begin{figure}\label{fig14}
\begin{center}
\includegraphics[scale=.65,angle=0]{clust2.pstex}
\caption{ The 
  two ($n=1$)-winding sectors are considered as belonging to a unique sector.   }
\end{center}
\end{figure}
For the  scaling (with the square of the radius of gyration) of the average area
enclosed by such   winding sectors, the numerical 
simulations give a result   similar  to what was obtained previously. 
The same conclusion holds  for  the ratio  $\lambda _+/\lambda _-$
which  remains practically  unchanged.
Considering now  the perimeter scalings,  for the external frontier  scaling
 the linear fit in figure 15 a) is $9.9 \; r_g^{4/3} $ meaning that the Hausdorff dimension remains  
  $d_H=4/3$.
Now, for the external+other windings and external+other windings+back-trackings frontiers scalings in figure 15 b) and figure 15 c) respectively,  linear
 fits with slopes slightly different from $4/3$ might be expected. However, a thorough analysis of
  the ratio of the frontiers  perimeters yields  the  fits 
       $10 \; r_g^{4/3} (\ln r_g)^{0.2} $ and  
  $15.5 \; r_g^{4/3} (\ln r_g)^{0.2} $. It follows that, up to logarithmic corrections, 
  the Hausdorff dimension  remains again unaffected to the value $d_H\approx 4/3$.
\begin{figure}\label{fig15}
\begin{center}
\includegraphics[scale=.3,angle=0]{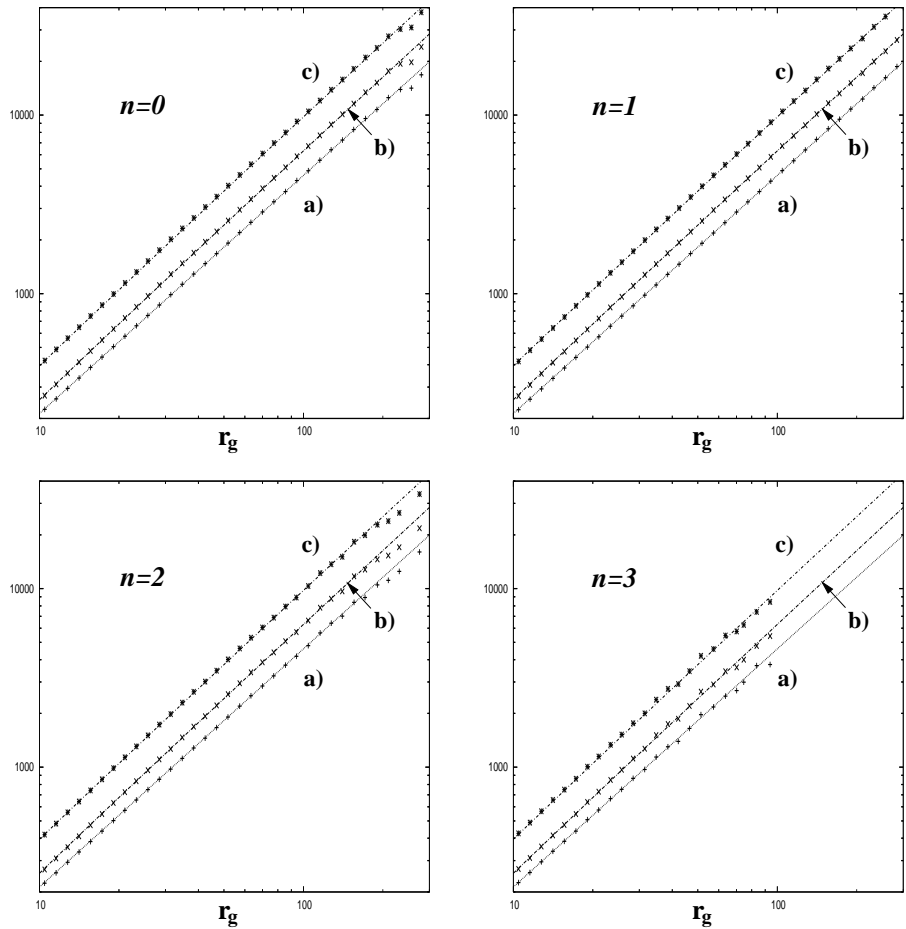}
\caption{numerical results for  the frontiers perimeters of the $n$-winding sectors ($n=0,1,2,3$);
 a)  external frontier; b) 
 external+other windings frontier; c) external+other windings+back-trackings frontier; the slope of the straight line
 a) is 4/3. For   b)  and  c), logarithmic corrections are included 
 (see the text).  }
\end{center}
\end{figure}
One concludes that the Hausdorff dimension of the winding sectors
inside the random walk,
 is, as for the random walk itself, $d_H=4/3$.

\section{Concluding remarks}\label{sec4}

Let us summarize the results obtained so far:

\begin{itemize}
\item we have discussed the correspondance between lattice random walk simulations and Brownian curves.
The correspondance  works well for global properties of the curves, as, for example, their radius of gyration, anisotropy  and   total arithmetic area. However,  the  arithmetic area of the $n$-winding sectors does not reproduce the path integral result for Brownian curves (eventhough, as already said,  it does for their sum, i.e. the total arithmetic area).
\item we have proposed a new definition for the  frontier  of a random walk and of a Brownian curve. 
It spans the $0$-winding sectors fjords connected to the sea. Contrary to the geometrical external frontier (the hull), this frontier has a coherent orientation.
\item we have numerically established that, for the random walk,    the Hausdorff dimension  of both its external (geometric) frontier and its  external+fjords (oriented) frontier (which does incursion inside the walk around the fjords) is, as for Brownian curves,  $d_H=4/3$.
On the other hand, the Hausdorff dimension of the external+fjords+lakes  (disconnected) frontier is $d_H\approx 7/4$, which indicates that the $0$-winding lakes deep inside the walk contribute in a major way to the fractal properties of the frontier. 
\item    we have numerically established that, for the random walk winding sectors, the Hausdorff dimension is robust at the value $d_H=4/3$. The scaling of the area of a winding sector with its radius of gyration  is also  very close to the scaling of the random walk itself.  
\end{itemize}

Let us  stress again that the scalings of  
the winding sectors  are not obvious  to interpret.   
  Still one has found on a 2d square lattice
strong numerical
 evidences which indicate  some deep similarities between the
random walks and their winding sectors -anisotropy,
 Hausdorff dimensions of the frontiers-, at least as far as the scalings are concerned.   
It would certainly be satisfying to understand more deeply these numerical
results.
Again, the similarity between the
 random walks and their winding sectors is especially puzzling.

The scaling dimension $4/3$ is known to be related via SLE to a conformal field
theory with central charge $0$
 (free fermions). For the scaling dimension $7/4$, the central charge
 is again $0$. However, the eventual link between those conformal
 models and closed random walks  frontiers or winding sectors
 frontiers is far from obvious. 
   Clearly, the $0$-winding sectors   play an essential role for
    increasing the Hausdorff dimension of the closed random walk  from
    $d_H=4/3$ to $d_H\approx 1.77$. On  the other hand, we have seen in Section  \ref{sec1}
  that the $0$-winding sector arithmetic area $<S_0>$ is overestimated   on the lattice. 
One can  expect that the above question is related to a
more fundamental  one: what happens, in the continuum limit, i.e. for
Brownian curves, to the winding
 sectors properties of random walks?

\vskip2cm

One of us (J.D.) acknowledges Thierry Jolicoeur  for his  help
in vectorizing the programs. We thank  Jesper Jacobsen for drawing our attention to 
 the question of $n$-winding sector  frontiers scaling and  for his participation at the early stage of 
 this work. We also thank Paul  Wiegmann for interesting
discussions.

\end{document}